\documentclass[preprint]{revtex4-1}
\usepackage[latin1]{inputenc}
\usepackage{amsmath,amssymb,amsbsy,amsfonts}
\usepackage{graphicx}
\usepackage{xcolor}
\usepackage{soul}

\begin{document}
	\title{Symmetry- and Input-Cluster Synchronization in Networks}
	\author{Abu Bakar Siddique$^a$,  Louis Pecora$^b$, Joe Hart$^c$, Francesco Sorrentino$^a$}
	\affiliation{$^a$ Department of Mechanical Engineering, University of New Mexico, Albuquerque, NM 87131, USA \\
    $^b$ Code 6343, Naval Research Laboratory, Washington, D.C. 20375, USA\\
    $^c$ Department of Physics, University of Maryland at College Park, MD 20742, USA}
	
    \begin{abstract}   
We study cluster synchronization in networks and show that the stability of all possible cluster synchronization patterns depends on a small set of Lyapunov exponents. Our approach can be applied to clusters corresponding to both orbital partitions of the network nodes (symmetry-cluster synchronization) and equitable partitions of the network nodes (input-cluster synchronization.) Our results are verified experimentally in networks of coupled opto-electronic oscillators. 
    \end{abstract}
    
    \maketitle


\section{Introduction} \label{sec: intro}
	Cluster synchronization (CS) in networks of coupled oscillators has been the subject of vast research efforts, see e.g., \cite{belykh2011mesoscale,ji2013cluster,pecora2014cluster,sorrentino2016complete,cho2017stable}. Recent work \cite{pecora2014cluster,sorrentino2016complete} has elucidated the relation between the symmetries of the network topology and the formation of clusters of synchronized dynamical units in the network. Reference \cite{pecora2014cluster} analyzed the formation and stability of synchronized clusters, where these clusters correspond to the orbits of the network symmetry group. Reference \cite{sorrentino2016complete} extended this study to different cluster synchronization patterns, where all of these are the result of symmetry breakings of the largest possible clusters predicted by the symmetry analysis. Reference \cite{sorrentino2016complete} also considered the formation of additional cluster synchronization patterns, not predicted by the symmetry analysis, for the case of networks whose connectivity is given in the form of a Laplacian matrix ($L$-networks). 

Here we consider the important problem of studying whether any pair of nodes (or subset of nodes) belonging to the same cluster synchronizes or not. Note that this is a more general problem than studying whether an entire cluster synchronizes or not. 
Comparing with the methods in \cite{schaub2016graph,cho2017stable}, 
our approach provides essential information that is key to addressing this problem. 
Moreover, different from the method in \cite{cho2017stable} our method can be applied \textcolor{black}{to explain the emergence of synchronization for nodes that are not related by symmetries, but receive the same total amounts of inputs from their neighboring nodes in different clusters. In what follows, we thus distinguish between \emph{symmetry-cluster synchronization} and \emph{input-cluster synchronization} and relate these two concepts to 
{ the network \emph{orbital partition} and \emph{equitable partition}, respectively. As we will see, the general rule is that nodes in the same cluster of the equitable partition can synchronize, while nodes in different clusters of the equitable partition cannot synchronize. Our main result is a method that is built on top of the symmetry analysis (orbital partition) and that allows us to determine whether any pair or subset of nodes belonging to a cluster of the equitable partition will synchronize or not.} }

%
\textcolor{black}{The paper is organized as follows.
In Sec. \ref{sec: partitions}, we introduce the orbital and equitable partitions of the nodes of a network. In Sec.\ \ref{sec: stability analysis}, we study the stability of the synchronous solution for nodes belonging to clusters corresponding to the orbital and equitable partitions.
In Sec.\ \ref{sec: cluster synchronization} 
we describe the stability of the orbital and equitable clusters {\color{black}and present an experimental verification of their stability using a network of optoelectronic oscillators \cite{hart2017experiments}}. 
Finally in Sec.\ \ref{sec: ET partition of A} we discuss the emergence of clusters not predicted by the symmetry analysis in the case of networks whose connectivity is given in the form of an adjacency matrix.
}
%

\section{Orbital and equitable partitions of the network nodes} \label{sec: partitions}

A network can be described by an adjacency matrix $A=\{A_{ij}\}$, whose entries 
	$A_{ij}=A_{ji}=1$, if node $j$ is connected to node $i$ and $A_{ij}=A_{ji} = 0$ otherwise. We call a so constructed network an $A$-network. We introduce the group $Aut$ of symmetries  of the network topology, where each symmetry permutes the network nodes but leaves the network topology unaltered. 
From knowledge of all the symmetries, the network can be partitioned into $M$ \emph{orbital} clusters \footnote{{\color{black} Here, with the term \emph{clusters} we indicate the subsets in which the set of the network nodes is partitioned. Throughout this paper we also refer to cluster synchronization as a dynamical phenomenon in which subsets of nodes synchronize on the same time evolution}}. Mathematically,  the orbit of a node $x$, $\mathcal{C}_o(x)$ is defined as the set of nodes $y$ such that $\exists \pi \in Aut: y=\pi*x$, where $\pi*x$ denotes the node $y$ that node $x$ is mapped to under the action of the symmetry $\pi$. 
	Under the action of all the network symmetries, the network nodes are partitioned into orbits formed of nodes that permute among one another.
	For example, the orbits of the network in Fig.\ 1a are $\{1,2,3,4\}$, $\{5,6,7,8\}$ and $\{9,10\}$.  


\begin{figure}[ht!]
	\centering
    \begin{tabular}{ccccc}
    	\includegraphics[width=0.275\textwidth]{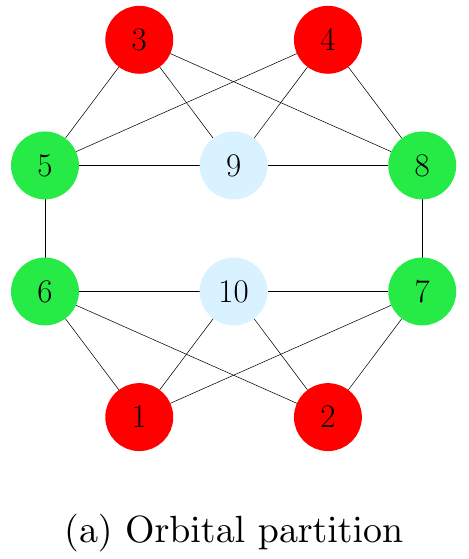}
        & \hspace {0.25cm} &
        \includegraphics[width=0.275\textwidth]{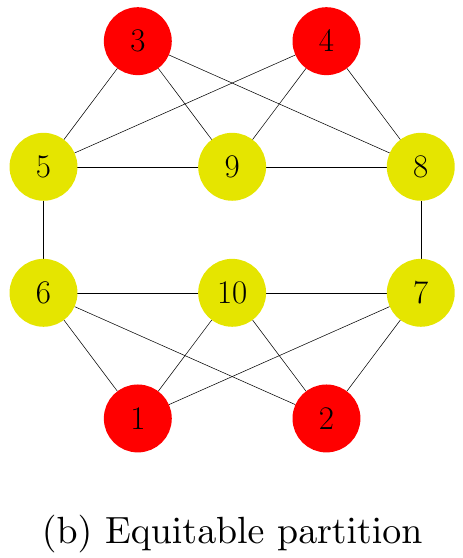} & \hspace {0.25cm} & \includegraphics[width=0.3\textwidth]{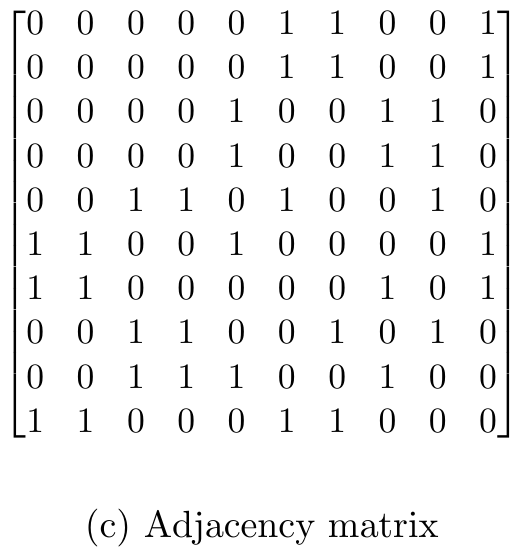}\\ [1ex]
\multicolumn{5}{c}{
\begin{tabular}{ccc}
\includegraphics[width=0.4\textwidth]{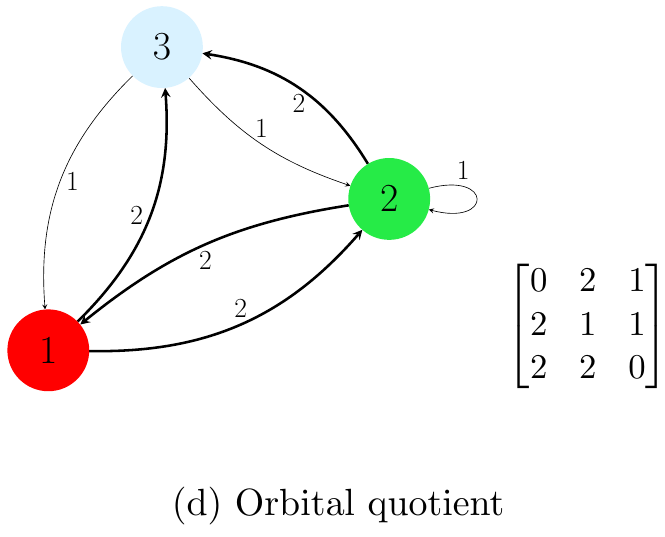} 
& \hspace {0.75cm} & \includegraphics[width=0.4\textwidth]{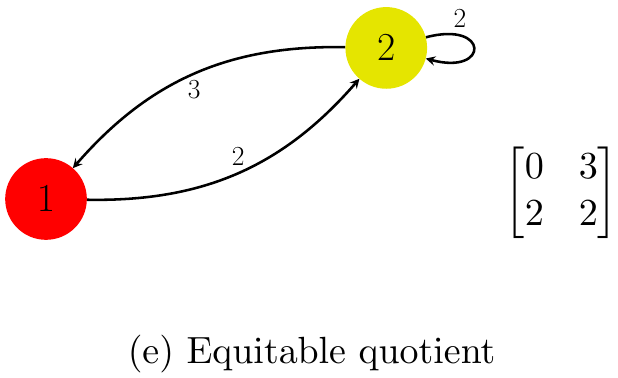}
\end{tabular}
}    

	\end{tabular}
	\caption{\textcolor{black}{Different representations of a} network: (a) shows 3 orbital clusters (red, green and blue) for the orbital partition and (b) shows 2 equitable clusters (red and yellow) for the equitable partition. (c) The network adjacency matrix $A$. (d) and (e) show the two quotient networks corresponding to the
partitions in (a) and (b), respectively. For each quotient network, the corresponding quotient matrix $Q$ is also shown.}	
	\label{fig: Network Partitions}	
\end{figure}

Another approach to partition the nodes of a network is based on an equitable partition \cite{schwenk1974computing}. In an equitable partition, any two clusters $\mathcal{C}_e^i$ and $\mathcal{C}_e^j$ of the partition are such that a node $x_i \in \mathcal{C}_e^i$ has exactly $n_{ij}$ neighbors in $\mathcal{C}_e^j$ regardless of the choice of $x_i$.  It is important to note that every orbital partition is also equitable but the converse is not true \cite{schaub2016graph}. Given a network, there may be several equitable partitions; out of these there is one equitable partition that is the coarsest, i.e., with the minimum number of clusters. A scalable numerical algorithm to obtain the coarsest equitable partition of a network is presented in \cite{belykh2011mesoscale}. In what follows we will refer to this as the equitable partition of a network and to the corresponding clusters as the equitable clusters. 
    
The condition for a CS pattern to be flow-invariant is that it corresponds to an equitable network partition,
	because in that case each node in each cluster receives the same number of \emph{inputs} from all the clusters \cite{schaub2016graph}.

In Fig.\ \ref{fig: Network Partitions} we show an $A$-network for which the orbital (Fig. \ref{fig: Network Partitions}a) and the equitable (Fig.\ \ref{fig: Network Partitions}b) partitions do not coincide. The associated adjacency matrix is shown in Fig.\ 1(c). 
Note that the green and blue clusters of Fig.\ \ref{fig: Network Partitions}a \emph{merge} to form the yellow cluster in Fig.\ \ref{fig: Network Partitions}b. All nodes of the yellow cluster are connected to two nodes of the red cluster and two other nodes of the yellow cluster. This makes it possible for all the nodes of the yellow cluster to synchronize. Similarly, all nodes of the red cluster are connected to three nodes of the yellow cluster, allowing all nodes of the red cluster to synchronize.

Given either an equitable or an orbital partition, we define the $M\times M$ quotient matrix $Q$ such that for each pair of clusters ($\mathcal{C}^u,\mathcal{C}^v$),
	\begin{equation}
	Q_{uv}=\sum_{j\in\mathcal{C}^v}A_{ij}, \qquad i\in\mathcal{C}^u,\quad u, v=1,2,\dots M,
	\label{eq: quotient matrix}
	\end{equation}
	where $M$ is the number of the clusters of the partition. The network in Fig.\ \ref{fig: Network Partitions} has $M=3$ orbital clusters and $M=2$ equitable clusters. The two quotient networks 
are shown in Fig.\ \ref{fig: Network Partitions}(d-e). As can be seen from Fig.\ \ref{fig: Network Partitions}, each cluster of the full network maps to a single node of the quotient networks. In what follows, when performing the stability analysis, we will linearize the full network equations about the desired CS solution corresponding to one of these two quotient networks.

\section{Stability Analysis} \label{sec: stability analysis}

A set of general dynamical equations to describe a system of coupled oscillators in discrete time is:
	\begin{equation}
		\pmb{\phi}_i\left[n+1\right]=\mathbf{F}(\pmb{\phi}_{i}\left[n\right]) + \sigma \sum_{j}^{N}A_{ij}\mathbf{H}(\pmb{\phi}_{j}\left[n\right]),
	\label{eq: dynamics discrete time}
	\end{equation}
	$i=1,2,\dots N$. 
    By grouping together all the nodes in the same clusters, Eqs.\ \eqref{eq: dynamics discrete time} become,
      \begin{equation}
		\pmb{\tilde \phi}_u\left[n+1\right]=\mathbf{F}(\pmb{\tilde \phi}_{u}\left[n\right]) + \sigma \sum_{v}^{M}Q_{uv}\mathbf{H}(\pmb{\tilde \phi}_{v}\left[n\right]),
		\label{eq: Quotient Equation}
	\end{equation}  
    $u=1,2,\dots M$. Note that Eqs.\ \eqref{eq: Quotient Equation} can be written for both the orbital and equitable network clusters. 
    
We consider here the dynamics of coupled opto-electronic oscillators \cite{williams2013experimental,pecora2014cluster,sorrentino2016complete,hart2017experiments} which is modeled by Eqs.\ \eqref{eq: dynamics discrete time}, where the scalar $\phi_i[n]$ is the phase of oscillator $i$ at time $n$, with $\mathbf{F}(\phi_i) = \beta I (\phi_i)$, $\mathbf{H}(\phi_i) = I(\phi_i)$ and $I(\phi_i) = \sin^2(\phi_i+\delta)$. Without loss of generality we set $\beta=3.5$ and $\delta=\pi/4$ and leave the coupling parameter $\sigma$ as a free parameter. 
		
\begin{figure}
    	\includegraphics[width=0.5\linewidth]{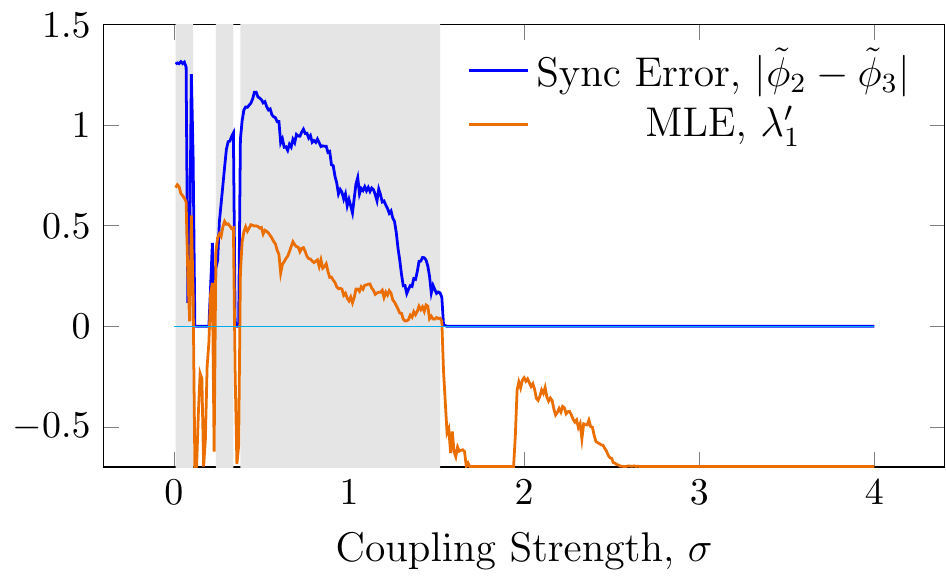} 
	\caption{ Synchronization Error and MLE vs. $\sigma$. The shaded region indicates the ranges of $\sigma$ for which the MLE, $\lambda_1'$ is positive. $\beta=3.5$ and $\delta=\pi/4$. }
	\label{fig: MLE}
\end{figure}

\subsection{Stability of the orbital and equitable quotient network dynamics}

An important preliminary step to proceed with our analysis that follows is to characterize stability of the two quotient networks in Fig.\ \ref{fig: Network Partitions}c and \ref{fig: Network Partitions}d. We linearize the orbital quotient network dynamics \eqref{eq: Quotient Equation} about the equitable quotient network state $\tilde \phi_2[n]=\tilde \phi_3[n]=\tilde \phi_s[n]$. 
We then study the time evolution of the perturbation $\tilde{\Delta}\phi=(\tilde \phi_2- \tilde \phi_3)$ (for more details see Appendix A), yielding,
\begin{equation}
\tilde{\Delta}\phi[n+1]=(\beta - \sigma)\sin (2 {\tilde \phi}_s[n]+ 2 \delta)\tilde{\Delta}\phi[n]
\label{eq: cluster stability}
\end{equation}
In Fig. \ref{fig: MLE}  we show the synchronization error $|\tilde \phi_2- \tilde \phi_3|$ and the maximum Lyapunov exponent (MLE) for Eq.\ \eqref{eq: cluster stability} as a function of $\sigma$.
{\color{black} Note that we are then characterizing not only the stability of a partition, but the stability of a partition that behaves in the particular way described by $\tilde \phi_s[n]$. 
Figs.\ 4, 5, 7 and 8 show MLE calculations over two different ranges of $\sigma$: shaded regions (non-shaded regions) correspond to ranges of $\sigma$ for which the MLE in Fig.\ \ref{fig: MLE} is positive and we linearize about the orbital quotient network dynamics (is negative and we linearize about the equitable quotient network dynamics). For the equitable quotient network dynamics, we numerically found that the same attractor is reached independent of the the initial conditions $\tilde \phi_1[0], \tilde \phi_2[0]$, when these are randomly chosen in the interval $(0, 2\pi)$.  The non shaded regions of Figs.\ 4, 5, 7, 8 indicate that the MLE's have been computed around this attractor. For the orbital quotient network dynamics, calculations were performed around the attractor 
to which Eq.\ (3) converges when it is evolved from $\tilde \phi_2[0] \simeq \tilde \phi_3[0]$.
The shaded regions of Figs.\ 4, 5, 7, 8 indicate that the MLE's have been computed around such an attractor.}
{\color{black}
It is also worth mentioning that it is possible to find a different asymptotic solution for $\tilde \phi_2[n]\neq\tilde \phi_3[n]$ with a completely different associated MLE. This however does not affect the results shown in what follows. }

{\color{black}
Moreover, we have studied the full dynamics of the orbital quotient network and of the equitable
quotient network to understand the relation between cluster synchronization and the system's behavior.
Interestingly, we have seen that the MLEs are sensitive to whether the asymptotic quotient network
behavior is a fixed point, a limit cycle or a strange attractor. The drops in the MLEs are in fact related
with a blue-sky catastrophe, a bifurcation in which the strange attractor on which the system evolves
is destroyed and is replaced by a simpler attractor (in our case, a period 2 limit cycle). On the other
hand, the smooth increases of the MLEs are related with a smooth complexification of the asymptotic
trajectory due to a Feigenbaum cascade (i.e., a sequence of period doubling bifurcations).
}
%


Next we analyze stability of all the valid CS states for Eqs.\ (2). Using the technique in \cite{pecora2014cluster} we compute the transformation matrix $T$ (non-normalized) of the network based on the irreducible representations (IRRs) of the symmetry group. A code that performs these operations and outputs the matrix $T$ is provided in \cite{aroncode2014git}. This transformation matrix $T$ maps from the node coordinate system into the IRR coordinate system \cite{pecora2014cluster}. 
Moreover, $T A T^{-1}=B$, where $B$ is a block diagonal matrix. 
The trivial representation (rows of the matrix $T$ are such that $T_{ij}=1$ if node $j$ is in cluster $i$ and $T_{ij}=0$ otherwise) corresponds to the synchronous state and we place those rows at the top of the matrix $T$, which means the first $M \times M$ block of the matrix $B$ corresponds to perturbations parallel to the synchronization manifold. The remaining \emph{transverse blocks} correspond to perturbations orthogonal to the synchronization manifold \cite{pecora2014cluster}. Following \cite{pecora2014cluster}, when studying stability of the CS state, a set of decoupled variational equations can be written, each one corresponding to a block of the matrix $B$ (see Appendix B). The transverse blocks are those responsible for stability of the CS solution. 

\begin{figure}
	\begin{tabular}{c}
      \includegraphics[height=0.4\linewidth]{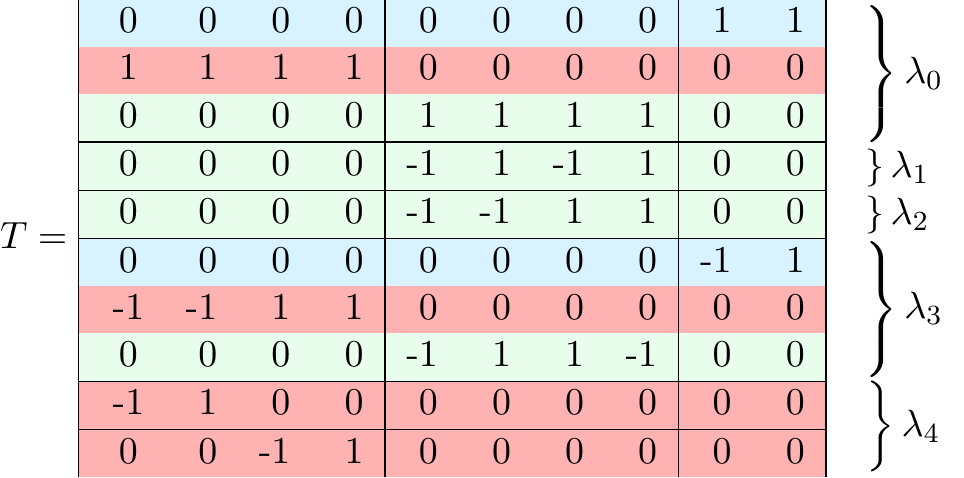} \\[2ex]
      \includegraphics[height=0.4\linewidth]{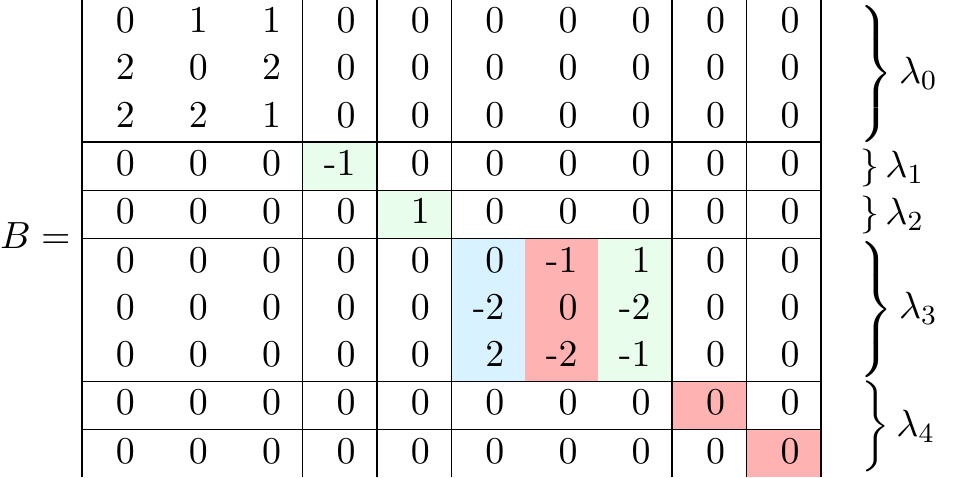}
    \end{tabular}
	\caption{The $T$ matrix based on the orbital partition and the corresponding $B$ matrix. Coloring is reminiscent of the colors used in Fig.\ \ref{fig: Network Partitions}a, namely pink rows correspond to the red cluster, light green rows to the green cluster and blue rows to the blue cluster.}
	\label{fig: orb T n B}
\end{figure}

\begin{figure}
	\def \figheight {0.22\linewidth}
	\begin{tabular}{ccc}
		\includegraphics[height=\figheight]{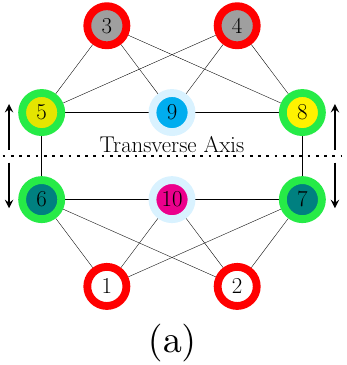}
		& \hspace {2cm} &
		\includegraphics[height=\figheight]{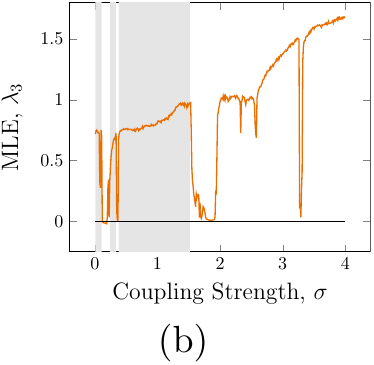} \\[2ex]
		\includegraphics[height=\figheight]{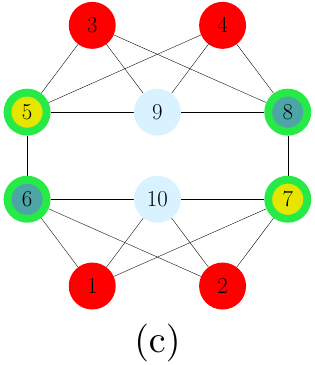}
		& \hspace {0.25cm} &
		\includegraphics[height=\figheight]{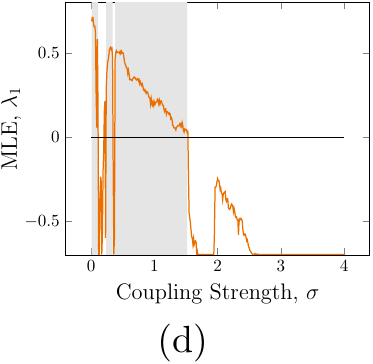} \\[2ex] 
		\includegraphics[height=\figheight]{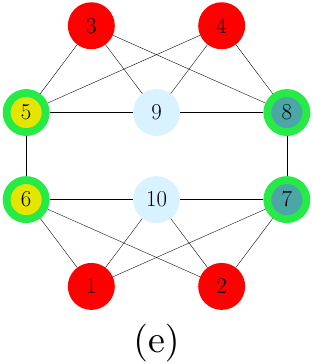}
		& \hspace {0.25cm} &
		\includegraphics[height=\figheight]{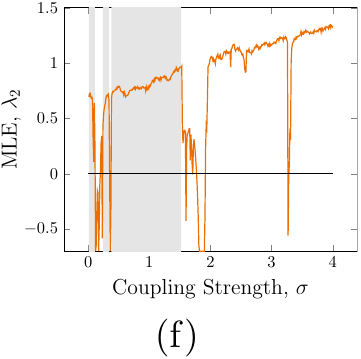} \\[2ex]
		\includegraphics[width=0.19\linewidth]{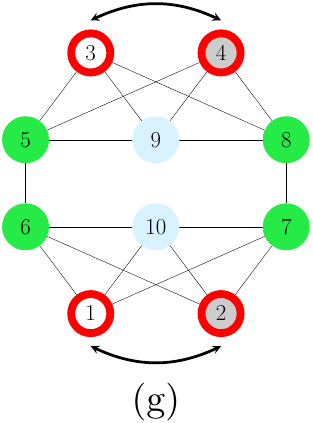}
		& \hspace {0.25cm} &
		\includegraphics[height=\figheight]{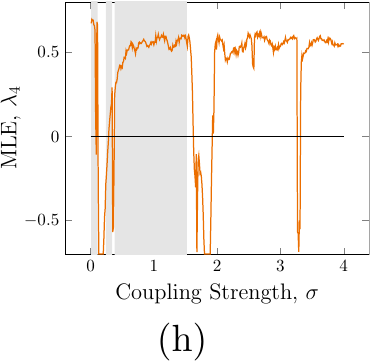}
	\end{tabular}
	\caption{\textcolor{black}{Plots a, c, e and g show different symmetry breakings of the network orbital partition. The inner circles inside each node describe such symmetry breakings. Two inner circles are colored the same (differently) if the nodes remain synchronized (do not remain synchronized) after the symmetry breaking.} Plots b, d, f and h show the MLE vs. the coupling strength $\sigma$ corresponding to the different symmetry breakings on the left (a, c, e, g).  shaded regions (non-shaded regions) indicate ranges of $\sigma$ for which the MLE in Fig.\ \ref{fig: MLE} is positive (negative.)}
	\label{fig: MLE_all}
\end{figure}

Referring to Fig.\ 3, it is important to note that each row of the matrix $T$ corresponds to a block of the matrix $B$ and so to a MLE. From these MLEs we can predict stability of every possible CS pattern. For the network in Fig.\ \ref{fig: Network Partitions}a, the first three rows of the transformation matrix $T$ correspond to the three orbital partition clusters (the trivial representation). The fourth row corresponds to the transverse block $\{-1\}$ of the matrix $B$ and the associated MLE is $\lambda_1$. The fifth row corresponds to the transverse block $\{+1\}$ of the matrix $B$ and the associated MLE is $\lambda_2$. Rows 6 to 8 correspond to the 3 by 3 transverse block of the matrix $B$, and the associated MLE is $\lambda_3$. This block determines stability about the \emph{transverse symmetry axis} in Fig.\ \ref{fig: MLE_all}a of all three clusters, which are therefore intertwined \cite{pecora2014cluster}. Rows 9 and 10 correspond to the transverse block $\{0\}$, with degeneracy 2, of the matrix $B$, and the associated MLE is $\lambda_4$. 

Figure \ref{fig: MLE_all} contains a graphical representation of how the rows of the matrix $T$ correspond to several symmetry breakings. The inner circles inside each node for the graphs on the left side of Fig. \ref{fig: MLE_all} describe such symmetry breakings. Two inner circles are colored the same (differently) if the nodes remain synchronized (do not remain synchronized) after the symmetry breaking. The condition for the symmetry breaking to occur is that the corresponding MLE is positive. The plots on the right hand side of Fig.\ \ref{fig: MLE_all} show the MLE associated with the symmetry breakings as a function of the free parameter $\sigma$. It is important to emphasize that the computation of each MLE is contingent on the quotient network dynamics, which is different in the case of the orbital and equitable partitions. In the plots on the rhs of Fig.\ \ref{fig: MLE_all}, the shaded areas (non-shaded areas) are for the MLEs being computed about the quotient network dynamics corresponding to the orbital (equitable) partition. 

Knowledge of these MLEs (i.e., the plots on the rhs of Fig.\ \ref{fig: MLE_all} for our example network) is all we need to compute stability of all the possible CS patterns. For example, we can assess whether synchronization of any pair of nodes in a cluster or synchronization of any cluster is stable or not. However, while the number of CS patterns can be extremely large \cite{sorrentino2016complete}, the number of these MLEs is upper bounded by  $N-M$, the number of transverse directions.  In the case in which stability of several CS patterns needs to be checked, this comes with an important computational advantage as the MLE calculations do not need to be repeated. As we will see, the matrix $T$ provides the connection between the MLEs and all the CS patterns.

Next we show that the transformation matrix $T$ contains essential information that can be used to assess stability of the synchronous state for different pairs of nodes (which can be generalized to any subset of nodes). We start from the assumption that nodes in different clusters of the orbital partition cannot synchronize \cite{sorrentino2016complete} (as we will see this assumption turns out being incorrect for the case in which the orbital partition and the equitable partition do not coincide.) Then we focus on understanding whether any two nodes in a cluster of the orbital partition synchronize or not. 
The matrix $T$ can be used as a \emph{table} to check whether any pair of nodes in the same cluster, say $i$ \textcolor{black}{and} $j$, synchronize or not based on the following steps,

\begin{itemize}
	\item Select columns $i$ and $j$ of the matrix $T$.
	\item Find all the rows $k$ of the matrix $T$ such that $T_{ki} \neq T_{kj}$. For each such row, consider the corresponding MLEs.
\end{itemize}

The condition for the two nodes $i$ and $j$ to synchronize is that all the Lyapunov exponents of the blocks corresponding to the rows $k$ for which $T_{ki} \neq T_{kj}$ are negative. Or analogously that the maximum of all those Lyapunov exponents is negative. For example, to compute the synchronization error between nodes $5$ and $7$, the rows of the matrix $T$ with different entries are $5$ and $8$, hence the MLEs are $\lambda_2$ and $\lambda_3$. Then the condition for nodes $5$ and $7$ to synchronize is that $\max(\lambda_2,\lambda_3)<0$. Following the same method, it is easy to see that the condition for the entire green cluster to synchronize is that $\max(\lambda_1,\lambda_2,\lambda_3)<0$.

Using this approach, we have  {\color{black}determined the stability} of different pairs of nodes belonging to the same cluster. {\color{black}The stability of the cluster states is verified by implementing the network from Fig. 1a in the experimental system described in detail in ref. \cite{hart2017experiments} as well as by direct simulation of Eq. \ref{eq: dynamics discrete time}.}
\begin{figure}[h!]
	\begin{tabular}{rcr}
		\includegraphics[height=0.365\linewidth]{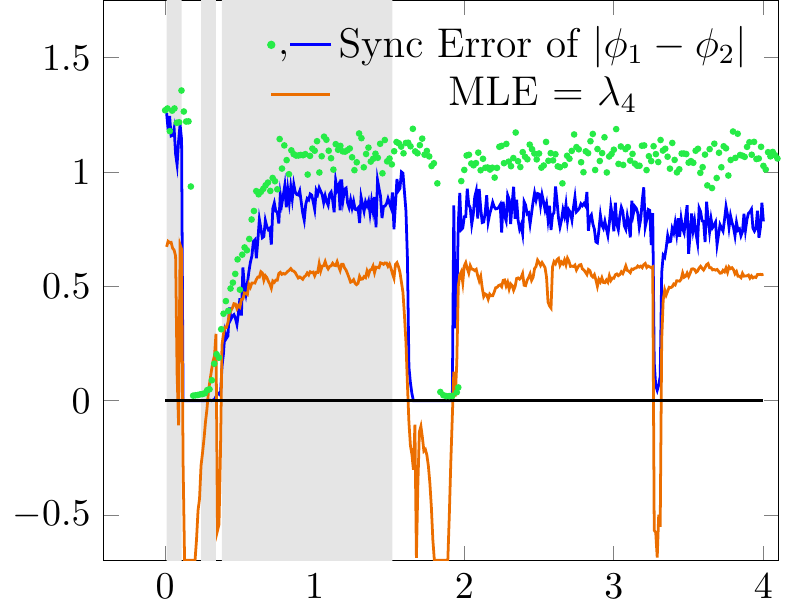} & \hspace{0.075\linewidth} &
		\includegraphics[height=0.365\linewidth]{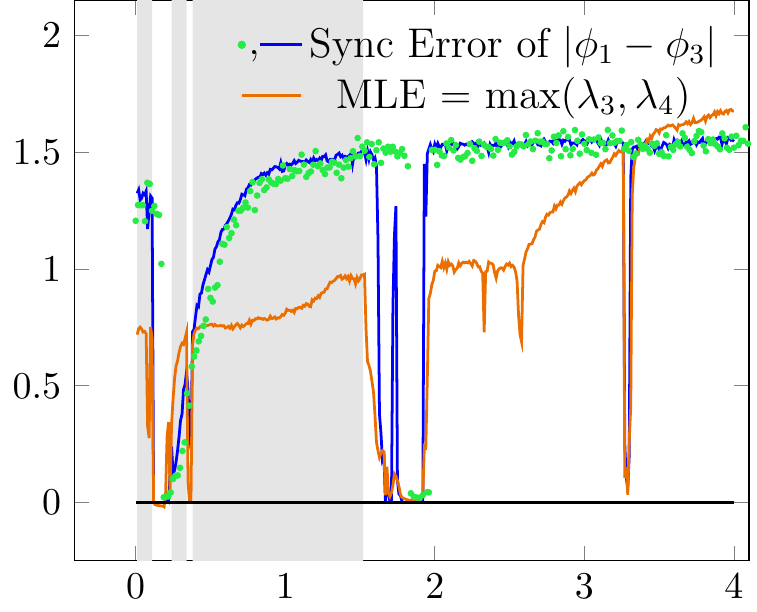} \\[2ex]
		
		\includegraphics[height=0.365\linewidth]{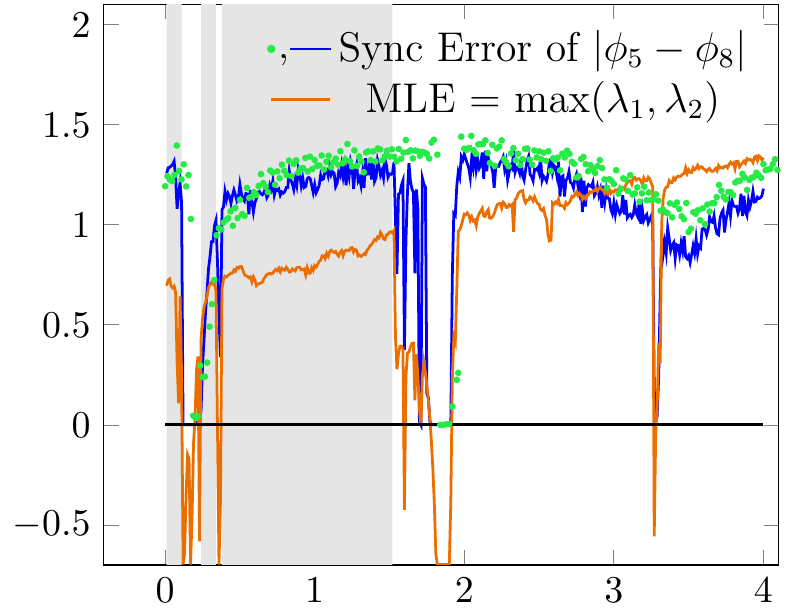} & \hspace{0.075\linewidth} &
		\includegraphics[height=0.365\linewidth]{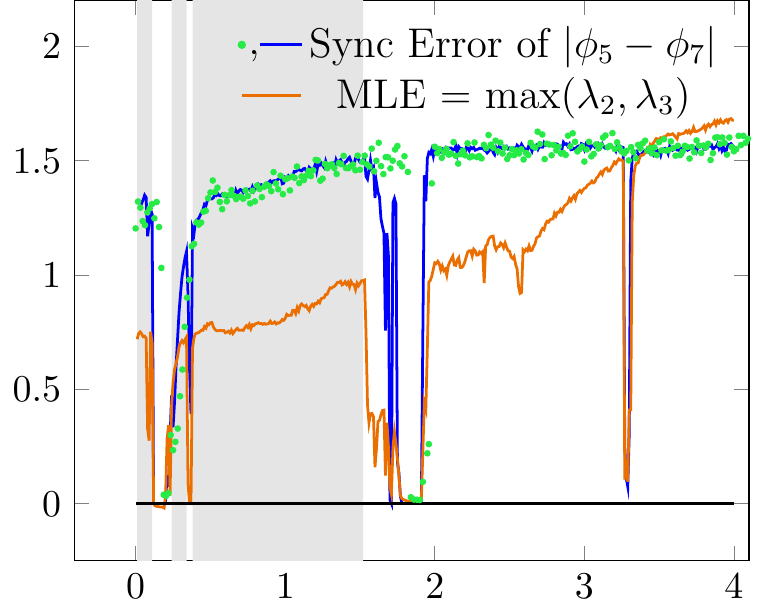} \\[2ex]
		
		\includegraphics[height=0.40\linewidth]{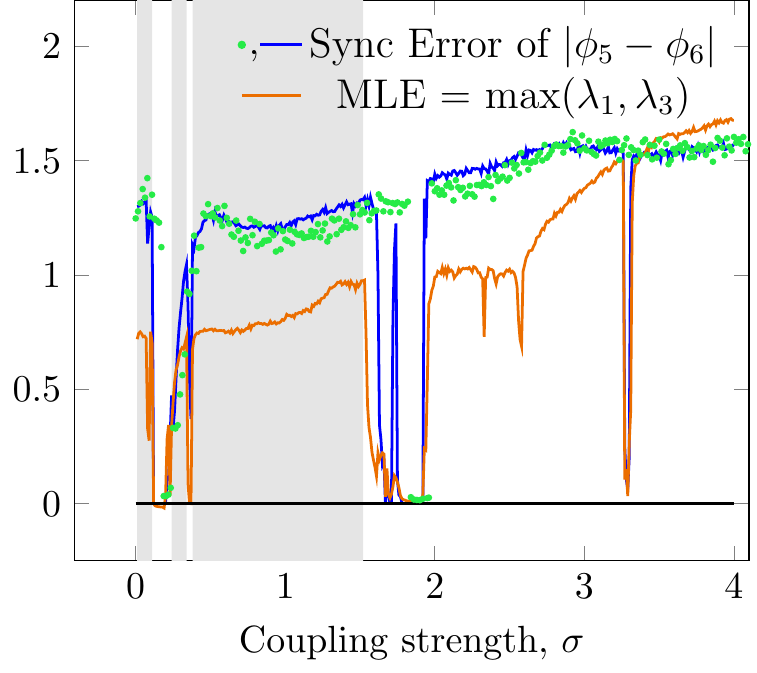} & \hspace{0.075\linewidth} &
		\includegraphics[height=0.40\linewidth]{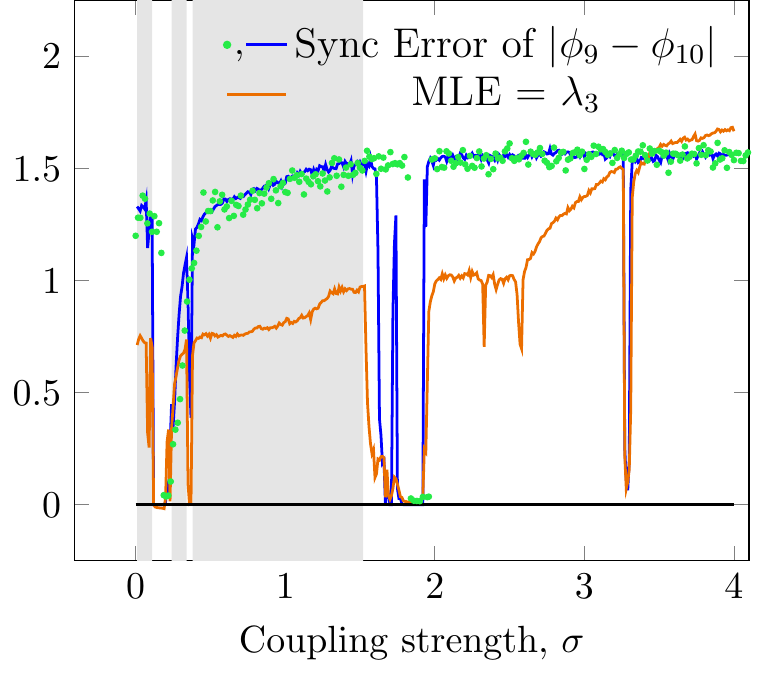}
	\end{tabular}
	\caption{Node-pairs synchronization errors (blue is from simulation and green is from experiment) and associated MLEs (in orange). Shaded regions indicates the range of $\sigma$ for which the MLE in Fig.\ 2 is positive. }
	\label{fig: Sync In Cluster}
\end{figure}
Figure \ref{fig: Sync In Cluster} shows synchronization errors {\color{black}from both experimental measurements and numerical simulations} and the associated MLEs for all the possible node pairs that can arise in the orbital partition. 
{\color{black} Because $\phi$ is defined modulo $2\pi$, we define synchronization error between two nodes $i$ and $j$ as $e\equiv\min(\langle|\phi_i-\phi_j|\rangle,\langle||\phi_i-\phi_j|-2\pi|\rangle)$, where $\langle\cdot\rangle$ indicates an average over time.

The experimental measurements validate the MLE predictions and the direct numerical simulations.} The shaded regions in the figure indicate the ranges of $\sigma$ for which the MLE in Fig.\ 2 is positive. 
\textcolor{black}{We also repeated our numerical simulations by adding a small amount of noise to the dynamics, for which we observed improved agreement with the $\sigma$-ranges of stable synchronization between experiment and simulation (not shown). 
}

\begin{figure}[h!]
	\begin{tabular}{c}
      \includegraphics[height=0.4\linewidth]{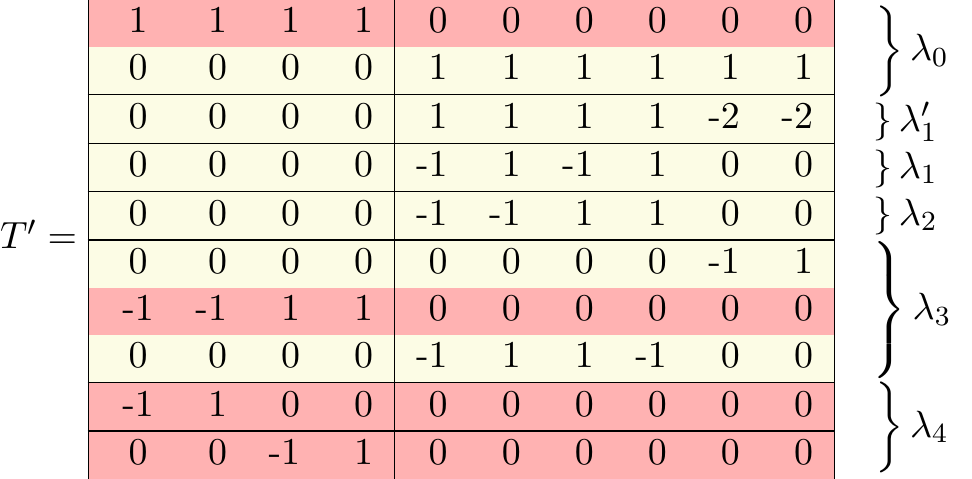} \\[2ex]
      \includegraphics[height=0.4\linewidth]{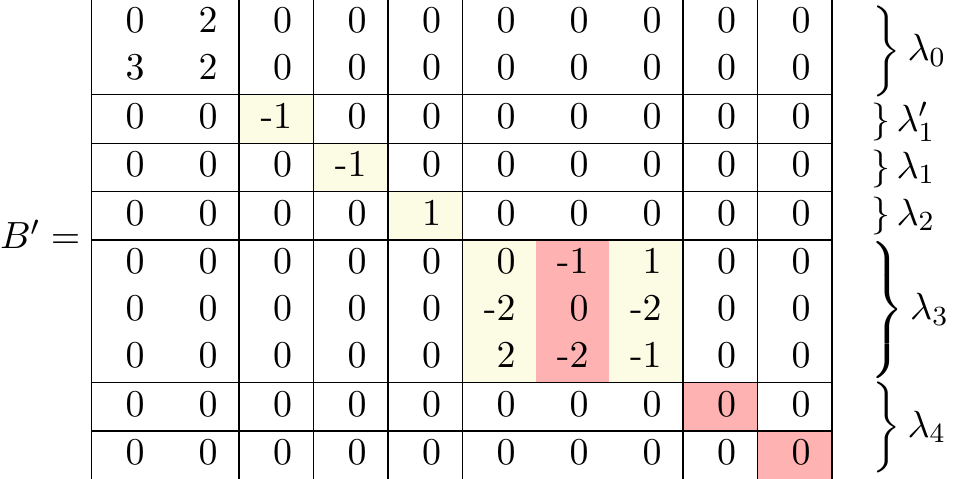}
	\end{tabular}
	\caption{The transformation matrix $T^{\prime}$ corresponding to the equitable partition and the associated block diagonal matrix $B^{\prime}$. Coloring is reminiscent of the colors used in Fig.\ \ref{fig: Network Partitions}b, namely pink rows correspond to the red cluster, light yellow rows to the yellow cluster.}
	\label{fig: ET T n B}
\end{figure}

We now consider the stability problem applied to the {equitable} partition. 
In Fig.\ \ref{fig: MLE} we observe that for certain ranges of $\sigma$, the equitable quotient network dynamics is stable, which suggests the possibility that pairs of nodes from the green and blue clusters can synchronize.
By following the method in \cite{sorrentino2016complete}, we can now construct a new transformation matrix $T^{\prime}$ that corresponds to the network equitable partition. Namely, we `merge' the green and blue clusters (Fig.\ \ref{fig: orb T n B}a), so to generate a new synchronous vector $\begin{bmatrix}
0 & 0 & 0 & 0 & 1 & 1 & 1 & 1 & 1 & 1 
\end{bmatrix}$ and an orthogonal vector $\begin{bmatrix}
0 & 0 & 0 & 0 & 1 & 1 & 1 & 1 & -2 & -2 
\end{bmatrix}$ in the plane determined by the first and third vectors/rows of the matrix $T$. In Fig.\ \ref{fig: ET T n B} we present the new transformation matrix $T^{\prime}$ and the corresponding matrix $B^{\prime}=T^{\prime}AT^{'-1}$.

As can be seen,  when the clusters are merged the dimension of the parallel block decreases by 1 and that of the transverse block increases by 1.  
A new transverse block is generated with associated MLE $\lambda_1'$. It is interesting to observe that in this case the \emph{new} MLE obeys the same equation as $\lambda_1$ in Fig. \ref{fig: MLE_all}d, hence $\lambda_1'=\lambda_1$.

\begin{figure}[h!]
	\begin{tabular}{lcr}
		\includegraphics[width=0.45\linewidth]{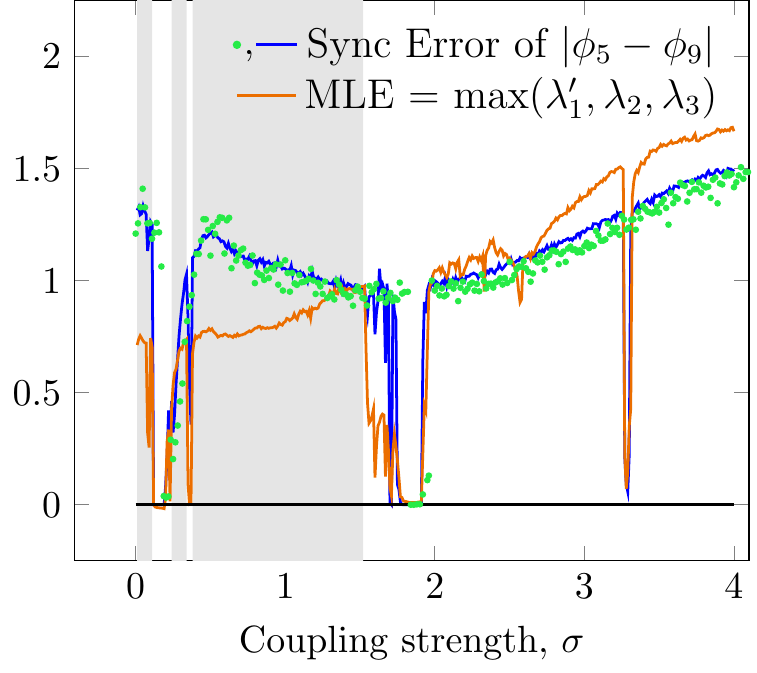} & 			      			\hspace{0.075\linewidth} &
		\includegraphics[width=0.45\linewidth]{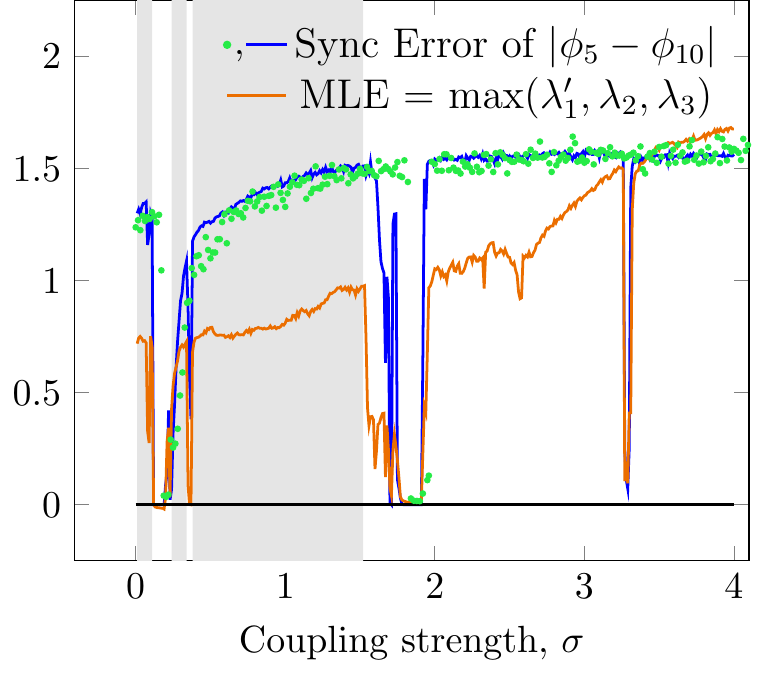}
	\end{tabular}
	\caption{Node-pairs synchronization errors (blue is from simulation and green is from experiment) and associated MLEs (in orange). Shaded regions indicates the range of $\sigma$ for which the MLE in Fig.\ 2 is positive. }
	\label{fig: sync_error_clusters}
\end{figure}

Since it is possible for nodes in the green and in the blue clusters to synchronize, then it makes sense to study whether pairs of nodes belonging to these clusters synchronize or not. In order to be able to perform this study, we follow the same procedure outlined before, but using the new \emph{table} $T^{\prime}$. Namely, for any pair of nodes $i,j$, we need to check whether the maximum of the MLEs corresponding to the rows $k$ of the matrix $T^{\prime}$ such that $T^{\prime}_{ki} \neq T^{\prime}_{kj}$ is positive or negative.
In Fig.\ 7 we show the synchronization error for the pairs of nodes (5, 9) and (5, 10) along with the associated MLEs. {\color{black} The experiments provide a validation of the theory.}


\section{Synchronization of the orbital and equitable clusters} \label{sec: cluster synchronization}

Here we describe how the symmetry breakings affect each one of the three clusters in Fig.\ \ref{fig: Network Partitions}a.
The red cluster can split into two pairs of nodes $\phi_1=\phi_2$ and $\phi_3= \phi_4$, which corresponds to a symmetry breaking occurring along the transverse symmetry axis (Fig. 4a). There are two other 
possible CS patterns of the red cluster: either $\phi_1=\phi_3$ and $\phi_2= \phi_4$ or $\phi_1=\phi_4$ and $\phi_2= \phi_3$.
The double transverse block $\{0\}$ describes stability of these two cluster synchronization patterns of the red cluster, so that depending on the associated MLE, they are either simultaneously stable or simultaneously unstable. The top and bottom bi-directional arrows in Fig.\ 4g represent the possibility of swapping nodes 3, 4 and 1, 2 in an arbitrary way.

Similarly, the green cluster can split into two pairs of nodes $\phi_5=\phi_8$ and $\phi_6=\phi_7$ which correspond to a symmetry breaking along the transverse symmetry axis (Fig. 4a). As can be seen in Fig.\ 4c and Fig.\ 4e, the green cluster can also break into two other possible pairs: either $\phi_5=\phi_7$ and $\phi_6=\phi_8$, which depends on the transverse block $\{-1\}$ or $\phi_5=\phi_6$ and $\phi_7=\phi_8$, which depends on the transverse block $\{+1\}$. Finally the blue cluster can break 
along the transverse symmetry axis in Fig.\ 4a, 
which depends on  the transverse $3\times 3$ block alone.

Similarly to what shown for pairs of nodes, the matrix $T$ ($T'$) can be used as a \emph{table} to check whether any subset of nodes from the same cluster synchronizes or not based on the following steps,

\begin{itemize}
\item Say the subset of nodes  from the same cluster of the orbital (equitable) partition is formed of nodes $i_1,i_2,...,i_p$. Select columns $i_1,i_2,...,i_p$ of the matrix $T$ ($T'$).
	\item Find all the rows $k$ of the matrix $T$ ($T'$) such that at least one of the entries  $T_{k{i_1}}, T_{kj{i_2}}, T_{kj{i_p}}$ ($T'_{k{i_1}}, T'_{kj{i_2}}, T'_{kj{i_p}}$) is different from the others. For such rows, consider the corresponding MLEs.
\end{itemize}

The condition for the subset of nodes $i_1,i_2,...,i_p$ to synchronize is that all the MLEs of the blocks corresponding to the rows $k$ for which at least one of the entries  $T_{k{i_1}}, T_{kj{i_2}}, T_{kj{i_p}}$ ($T'_{k{i_1}}, T'_{kj{i_2}}, T'_{kj{i_p}}$) is different from the others, are negative. Or analogously that the maximum of all those Lyapunov exponents is negative. Obviously, this approach applies also to the case that the subset of nodes coincides with the whole cluster.

\begin{figure*}[h!]
	\begin{tabular}{rcr}
		\includegraphics[width=0.4\linewidth]{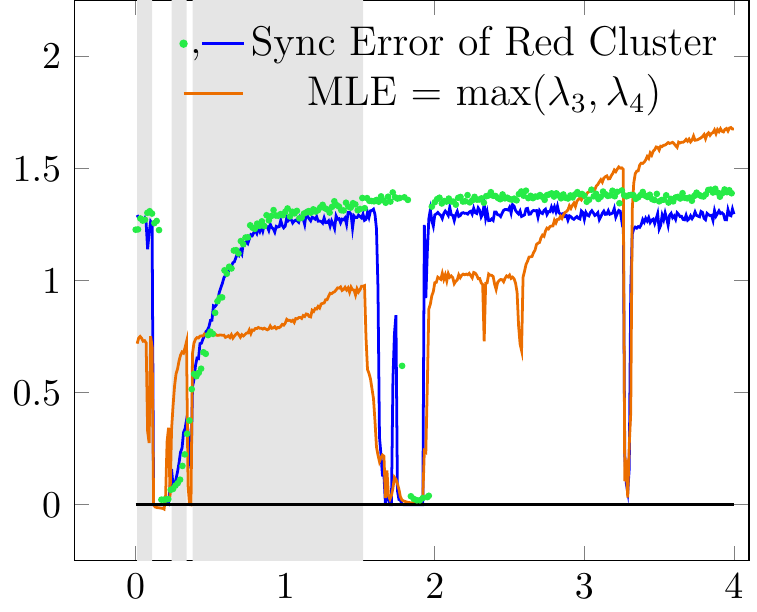} & \hspace{0.075\linewidth} &
		\includegraphics[width=0.4\linewidth]{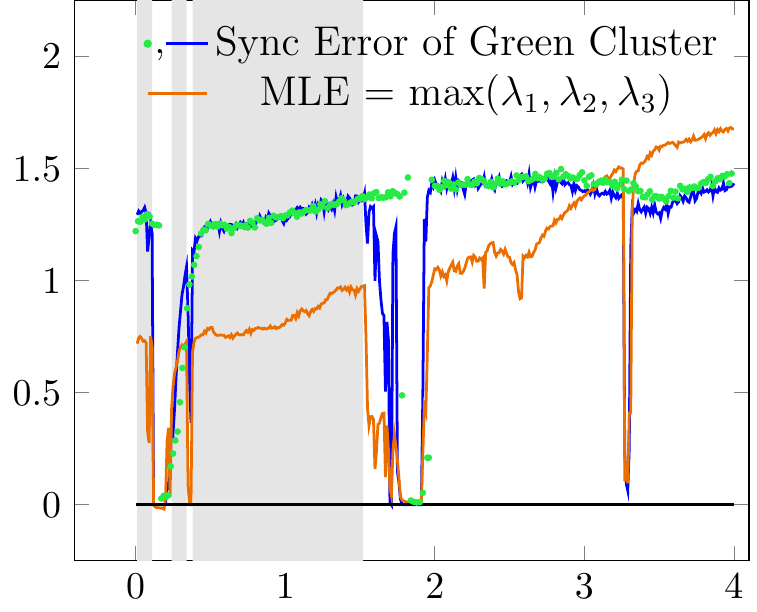} \\[2ex]
		
		\includegraphics[width=0.4\linewidth]{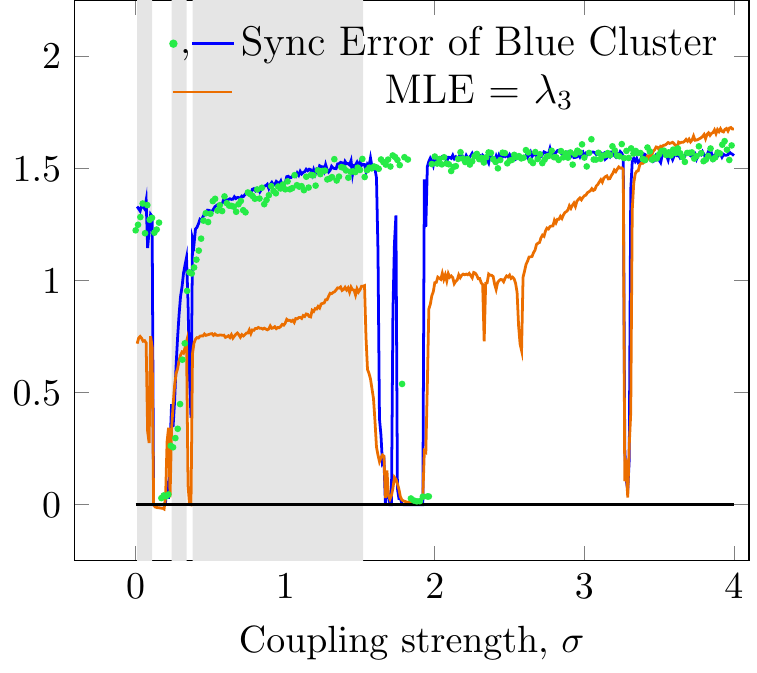} & \hspace{0.1\textwidth} &
		\includegraphics[width=0.4\linewidth]{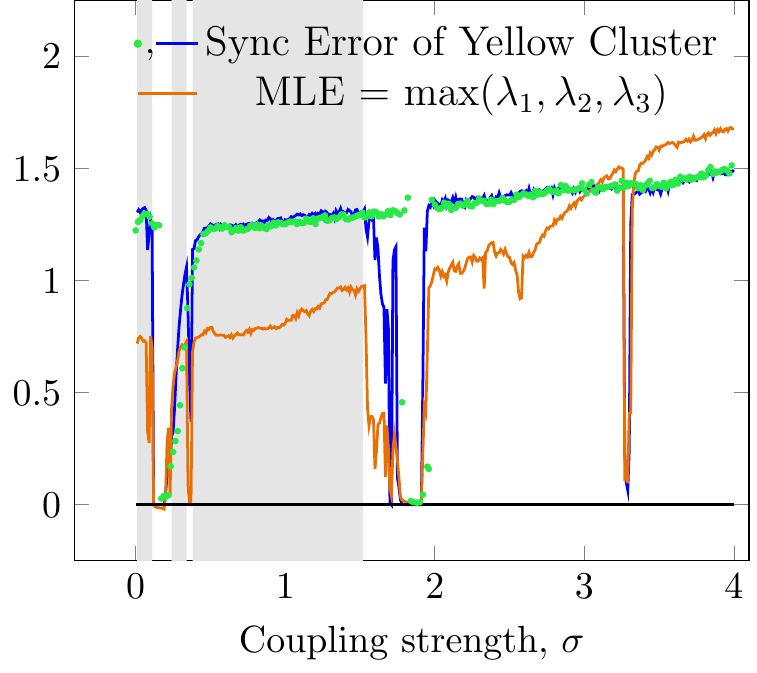}
	\end{tabular}
	\caption{Node-pairs synchronization errors (blue is from simulation and green is from experiment) and associated MLEs (in orange). Shaded regions indicates the range of $\sigma$ for which the MLE in Fig.\ 2 is positive.}
	 \label{fig: cluster sync all nodes}
\end{figure*}

Figure \ref{fig: cluster sync all nodes} shows the cluster synchronization errors, \textcolor{black}{ computed as the averaged pairwise synchronization error over all the possible pairs inside a cluster}, and associated MLEs for the red cluster, green cluster, blue cluster, and yellow cluster, and again we see agreement between our stability predictions and the synchronization errors. The shaded regions in the figure indicate the ranges of $\sigma$ for which the MLE in Fig.\ 2 is positive. 

\section{Equitable partitions of $A$-networks} \label{sec: ET partition of A}

Different from what was previously reported \cite{sorrentino2016complete}, here we have shown that the formation of cluster synchronous patterns not predicted by the symmetry analysis is also possible in the case of networks whose connectivity is given in the form of an adjacency matrix ($A$-networks). 

Based on previous work from some of us \cite{pecora2014cluster}, the network symmetries can be used to partition the nodes of an $A$-network into clusters corresponding to the orbits of the symmetry group: nodes in the same cluster can synchronize, while nodes in different clusters cannot synchronize. This statement does not hold in the case that the clusters predicted by the symmetry analysis do not coincide with those predicted based on the inputs each node receives \cite{belykh2011mesoscale,schaub2016graph}. It is indeed possible that two (or more) nodes may synchronize if they receive the same inputs from their neighboring nodes, even when they are not related by a symmetry relation. When this happens, one should instead consider the equitable partition which corresponds to a refinement of the orbital partition provided by the symmetry analysis \cite{belykh2011mesoscale,schaub2016graph}. We can then conclude that \emph{nodes in the same cluster of the equitable partition can synchronize, while nodes in different clusters of the equitable partition cannot synchronize}. Then the question arises whether any two nodes belonging to the same cluster of the equitable partition will synchronize or not. In this paper, we have proposed a method that provides an answer to this question. This method allows us to determine whether any pair of nodes belonging to a cluster of the equitable partition will synchronize or not. We have shown that though the symmetry analysis cannot be directly applied to analyze stability of these patterns, it can be conveniently extended to address this problem.

\section{Conclusions}

In this paper we have proposed a unified approach to determine whether any subset of nodes belonging to a cluster of the \emph{equitable} partition synchronizes or not. 
One of the most useful quantities to come out of the computational group theory analysis of a network for orbital and equitable 
clusters is the transformation matrix $T$. Not only does the matrix $T$ block diagonalize the coupling matrix, but the rows of $T$ are the bases of the IRR of the network symmetry group, so that each row shows the directions of the perturbations associated with that IRR. These are the directions each cluster's nodes will deviate in a symmetry breaking bifurcation. This is essential knowledge for obtaining normal forms for the bifurcation analysis and for identifying the new clusters created.  Note, this holds for 
equitable clusters and Laplacian clusters \cite{sorrentino2016complete}, too. Our results indicate a tradeoff between computational complexity \cite{cho2017stable} and completeness of information in addressing the problem of CS stability in networks, in that lower
computational complexity may result in less information about the
stability of the symmetry breakings.

\appendix
\section{Stability of the two Quotient Networks Dynamics}

In order to study stability of the full network dynamics, we first need to characterize stability of the two quotient networks in Fig.\ \ref{fig: Network Partitions}d and \ref{fig: Network Partitions}e. We write down equation \eqref{eq: Quotient Equation} for the quotient network in Fig. \ref{fig: Network Partitions}d,
\begin{equation}
\begin{split}
\mathbf{\tilde x}_{1}[n+1]=& \textbf{F}(\mathbf{\tilde x}_{1}[n]) + \sigma \left(2 \textbf{H}(\mathbf{\tilde x}_{2}[n])+\textbf{H}(\mathbf{\tilde x}_{3}[n]) \right) \\
\mathbf{\tilde x}_{2}[n+1]=& \textbf{F}(\mathbf{\tilde x}_{2}[n]) + \sigma \left(2\textbf{H}(\mathbf{\tilde x}_{1}[n]) + \textbf{H}(\mathbf{\tilde x}_{2}[n])+\textbf{H}(\mathbf{\tilde x}_{3}[n]) \right) \\
\mathbf{\tilde x}_{3}[n+1]=& \textbf{F}(\mathbf{\tilde x}_{3}[n]) + \sigma \left(2 \textbf{H}(\mathbf{\tilde x}_{1}[n])+ 2\textbf{H}(\mathbf{\tilde x}_{2}[n]) \right)
\label{eq: dynamics continuous time expanded}
\end{split}
\end{equation}

We are interested in studying stability of the equitable quotient network dynamics. The reason is that in order to analyze the CS of the full network, we are going to linearize the full network equations about either the quotient network dynamics for the equitable partition or for the orbital partition. Thus we linearize Eq. \eqref{eq: dynamics continuous time expanded} about the state ${\mathbf{\tilde x}}_s={ \mathbf{\tilde x}}_2={\mathbf{\tilde x}}_3$ corresponding to the equitable partition quotient network dynamics, yielding

\begin{subequations}\label{eq: linearized continuous time equation}
	\begin{align}
	\delta \mathbf{\tilde x}_{1}[n+1] =& D\textbf{F}( \mathbf{\tilde x}_{1})\delta \mathbf{\tilde x}_1[n] + \sigma D\textbf{H}(\mathbf{\tilde x}_{s})\left[2 \delta \mathbf{\tilde x}_2[n]+\delta \mathbf{\tilde x}_3[n] \right] \\
    \delta  \mathbf{\tilde x}_{2}[n+1] =& D\textbf{F}(\mathbf{\tilde x}_{s})\delta \mathbf{\tilde x}_2[n] + 2\sigma D\textbf{H}(\mathbf{\tilde x}_{1})\delta  \mathbf{\tilde x}_1[n] + \sigma D\textbf{H}(\mathbf{\tilde x}_{s}) \left[\delta \mathbf{\tilde x}_2[n]+\delta \mathbf{\tilde x}_3[n] \right]   \\
	\delta \mathbf{\tilde x}_{3}[n+1] =& D\textbf{F}(\mathbf{\tilde x}_{s})\delta \mathbf{\tilde x}_3[n] + \sigma \left[2 D\textbf{H}(\mathbf{\tilde x}_{1})\delta \mathbf{\tilde x}_1[n]+ 2D\textbf{H}(\mathbf{\tilde x}_{s})\delta \mathbf{\tilde x}_2[n] \right]
	\end{align}
\end{subequations}

We define ${\Delta}\mathbf{\tilde x}[n]=(\delta \mathbf{\tilde x}_2[n]-\delta \mathbf{\tilde x}_3[n])$ and we write,
\begin{equation}
{\Delta}\mathbf{\tilde x}_{2}[n+1]= \left[D\textbf{F}(\mathbf{\tilde x}_{s}) - \sigma  D\textbf{H}(\mathbf{\tilde x}_{s})\right] \Delta \mathbf{\tilde x}_2[n],
\label{eq: sync continuous time}
\end{equation}	
which is the same as Eq.\ (4).

We note that Eq.\ \eqref{eq: sync continuous time} does not depend on $\mathbf{\tilde x}_1$, which leads to the general rule for merging two clusters or quotient nodes:  both clusters or quotient nodes must receive exactly the same coupling from each other cluster or quotient node in the orbital partition. 
Note that in the orbital partition quotient network there is only one node for each cluster. Then a condition for two nodes of the orbital quotient network to merge is that they receive exactly the same coupling from each other node of the quotient network. 

Based on this analysis, if for a certain choice of parameters the quotient network for the equitable partition is found to be unstable, one would not linearize about this state. Conversely, if it is found to be stable, we know that linearizing about the orbital network partition may actually result in linearizing about the equitable network partition (depending on the choice of the initial conditions). 

\section{Decoupling the variational equations}

Consider the orbital partition of the network nodes in $M$ clusters,  ${\mathcal C}_m$, $m=1,...,M$, with synchronous motion ${\bf s}_m[n]$. Then we can write the original vectorial variational equations about the synchronized solutions in the node coordinates,

\begin{equation}\label{linv}
\delta {{\bf x}}[n+1]=\left[\sum_{m=1}^{M} E^{(m)} \otimes D{\bf F}({\bf s}_m[n] +  \sigma A \sum_{m=1}^{M} E^{(m)} \otimes D{\bf H}({\bf s}_m[n]) \right] \delta {{\bf x}}[n],
\end{equation}
where the $Nn$-dimensional vector $\delta {\bf x}[n]=[\delta {\bf x}_1[n]^T,\delta {\bf x}_2[n]^T,...,\delta {\bf x}_N[n]^T]^T$ and $E^{(m)}$ is an $N$-dimensional diagonal matrix such that
\begin{align}
E^{(m)}_{ii}=\left\{ \begin{array} {ccc} {1,} \quad \mbox{if} \quad {i \in {\mathcal C}_m,} \\ {0,}  \quad \mbox{otherwise,}  \end{array} \right.
\end{align}
$i=1,...,N$.
Note that $\sum_{m=1}^{M} E^{(m)}=I_N$, where $I_N$ is the $N$-dimensional identity matrix.

Applying $T$ to Eq.~(\ref {linv}) we arrive at the variational matrix equation shown in Eq.~(\ref {eq:2}),where ${\pmb \eta}(n)= T \otimes I_n \, \delta {\bf x}(n)$, $J^{(m)}$ is the transformed $E^{(m)}$, and $B$ is the block diagonalization of the coupling matrix $A$,
\begin{equation} \label{eq:2}
{{\pmb \eta}}[n+1]=\left[\sum_{m=1}^{M} J^{(m)} \otimes D{\bf F}({\bf s}_m[n] +  \sigma B\otimes I_n \sum_{m=1}^{M} J^{(m)} \otimes D{\bf H}({\bf s}_m[n]) \right] {\pmb \eta}[n],
\end{equation}
where we have linearized about synchronized cluster states $\{{\bf s}_1,...,{\bf s}_M\}$, ${\pmb \eta}[n]$ is the vector of variations of all nodes transformed to the IRR coordinates and $D{\bf F}$ and $D{\bf H}$ are the Jacobians of the nodes' vector field and coupling function, respectively.  

We can write the block diagonal $B$ as a direct sum $\bigoplus_{l=1}^L I_{d^{(l)}} \otimes B^{l}$, where $B^{l}$ is a  $p_l \times p_l$ matrix with $p_l=$ the multiplicity of the $l$th IRR in the permutation representation, $L=$ the number of IRRs present, and $d^{(l)}=$ the dimension of the $l$th IRR, so that $\sum_{l=1}^L d^{(l)} p_l=N$ \cite{pecora2014cluster}. The trivial representation ($l=1$) which is associated with the motion in the synchronization manifold has $p_1={M}$. The remaining transverse blocks all decouple in independent variational equations that determine stability of the $CS$ solution.

Acknowledgements. The authors are grateful to Raj Roy, Tom Murphy, and Fabio Della Rossa for insightful conversations.   

\end{document}